\documentclass[preprint,12pt]{elsarticle}
\graphicspath{ {./figures/} }
\usepackage{hyperref}
\hypersetup{hidelinks,colorlinks=true,linkcolor=red,citecolor=blue,urlcolor=black}
\usepackage{float}
\usepackage{verbatim} 
\usepackage{apalike}
\restylefloat{figure}
\floatstyle{plaintop} 
\restylefloat{table}

\journal{Journal of Informetrics}

\usepackage{natbib}
\biboptions{authoryear}

\begin{document}
\begin{frontmatter}

\title{Is a team only as strong as its weakest link? Quantifying the short-board effect with AI Agents}

\author[label1]{Xin Xu  \corref{cor1}}

\author[label2,label3]{Jiu Zhang \corref{cor1}}

\author[label1]{Xiao-Ling Lei}

\author[label4]{Xiong-Fei Jiang \corref{cor2}}
\ead{jiangxiongfei@nbufe.edu.cn}

\author[label5]{Long Xiong \corref{cor2}}
\ead{xionglong@zju.edu.cn}

\cortext[cor1]{These authors contributed equally to this work.}
\cortext[cor2]{Corresponding authors.}
\address[label1]{School of Business and Tourism Management, Yunnan University, Kunming, 650091, China}
\address[label2]{School of Digital Economy and Management, Fuyao University of Science and Technology, Fuzhou, 350109, China}
\address[label3]{Digital  Governance Laboratory, Fuyao University of Science and Technology, Fuzhou, 350109, China}
\address[label4]{College of Finance and Information, Ningbo University of Finance and Economics, Ningbo 315175, China}
\address[label5]{School of Physics and Astronomy, Yunnan University, Kunming, 650091, China }

\begin{abstract}
	The short-board effect, analogous to Liebig's Law of the Minimum, postulates that the collective performance of a team is constrained by its weakest component.
	This principle has profound implications for the optimization of collaboration in a variety of contexts, including management, education, and organizational structures.
	Despite its theoretical significance, empirical validation remains elusive due to challenges of assessing individual capabilities, controlling real-world variables, and data biases towards successful outcomes, as well as high employee turnover.
	To address this absence of knowledge, we employ multi-agents driven by large language models to simulate a teamwork with standard operating procedure, revealing the relationship between individual capability and collective team performance.
	In homogeneous team configurations, three capability regimes are observed, particularly the Sisyphus predicament state at the critical capability threshold characterized by extensive ineffective efforts and pseudo-high efficiency.
	Furthermore, with a single weak link quantifying  the short-board effect, we highlight different impacts across core and non-core members on the team performance.
	More importantly, when the team exhibits multiple weak links, a cumulative product effect emerges, demonstrating that team performance is shaped by the aggregated impact of all weaknesses rather than the weakest link solely.
	This suggests that mitigation strategies should extend beyond the remediation of individual weak links.
	These findings rigorously elaborate the short-board theory and provide actionable insights to optimize team management, organizational operations, and supply chain resilience.
\end{abstract}


\end{frontmatter}
\section{Introduction}
\label{introduction}

A team is only as strong as its weakest link, a timeless adage that underscores the critical role of the least capable element in determining collective performance.
This principle finds a direct parallel in Liebig's Law of the Minimum \citep{von1840organische}, originally proposed in agricultural chemistry to describe how plant growth is limited by the scarcest resource.
Then this idea is vividly described by the "barrel effect" where the shortest board determines the barrel's capacity, i.e., the term of "short-board effect".
Over time, the theory has been extensively applied beyond ecology, to fields such as management, economics, education, and other related disciplines, to explain how weaknesses within a team can diminish overall outcomes.
As an example, even when the same lead surgeon performs an identical operation, different communication protocols or resource availability of medical centers can significantly affect outcomes, with any weak link potentially resulting in failure \citep{dias2025team,kanamori2025implementation}.
These interdisciplinary applications highlight the need for rigorous empirical studies on the short-board effect to provide actionable insights for enhancing team efficacy.

Despite being a well-known theory, it is typically discussed in a qualitative manner, and its effectiveness remains to be verified.
Empirical research is relatively challenging for numerous reasons.
For example, in team collaboration, accurately assessing abilities of each team member presents significant difficulties and may involve workplace discrimination;
since team collaboration in real-world scenarios is influenced by many factors \citep{edmondson1999psychological,lia2025multi}, it is impossible to study the impact of individual weak link through controlled experiments;
most observable projects are successful ones, as failed projects are typically terminated early, resulting in data that are inherently biased toward successful samples and making it extremely difficult to obtain comprehensive datasets;
and in practice, high employee turnover and frequent team reorganizations lead to fragmented data.
Therefore, a systematic and quantitative examination of the short-board effect would not only rigorously test this theory, but also provide important insights for improving team collaboration management.
Nevertheless, the theoretical landscape in this domain is lacking.

Large language model (LLM), trained on vast amounts of textual data, have recently demonstrated exceptional capabilities in a variety of domains, including scientific writing, question answering, programming, etc \citep{simon2024language,kung2023performance,singhal2023large,guo2023can,sun2024scieval,li2025large,guo2025deepseek,capraro2024language,ma2025urban}.
When further guided by domain-specific prompts, LLMs have been shown to emulate the roles of scientists, project managers, engineers, trading agents, or physicians \citep{zhao2026ai,xi2025rise,guo2024large}, often achieving those of near-expert practitioners.
Of particular interest is the emerging paradigm of multi-agents systems, where multiple LLM-driven agents collaborate to solve complex tasks,
due to the growing demand for interdisciplinary expertise in contemporary scientific and engineering problems \citep{park2023generative,durante2024agentaisurveyinghorizons,park2026self,nouri2026agentic,yan2026large}.
Such approaches have already shown promise in fields including nanobodies \citep{swanson2025virtual}, software development \citep{hoffmann2024generative}, and financial analysis \citep{li2023tradinggpt}.
Beyond these practical applications, LLM-based multi-agent systems also serve as a powerful experimental platform to investigate fundamental questions in psychology and management science \citep{hagendorff2024deception,cui2025large,andrieux2024ethical,oswick2024generative,yan2026large},
thereby offering new opportunities for studying human behaviors such as decision-making and team collaboration \citep{hua2023war,lu2024llms,zimmaro2024emergence,guo2024large,ben25does,cheung25large}.
Exploring the capability boundaries and inherent characteristics of LLM-based agents collaboration provides valuable guidance for addressing real-world project management challenges,
and lays a crucial foundation for advancing multi-agents system applications.

Due to the variations in parameter architectures, LLMs manifest considerable disparities in problem-solving capabilities \citep{chiang2024chatbot}.
Consequently, employing different LLMs to drive agents enables the simulation of team members with varying competencies on the team collaboration.
By leveraging this characteristic, it is possible to overcome the challenge of controlling variables to assess individual contributions in real-world scenarios,
thereby allowing for quantitative analysis of the short-board effect in the collaborative workflow.
In this study, a virtual software development team driven by LLM is utilized.
By configuring teams with homogeneous and heterogeneous configurations, we investigate the relationship between the team performance and the capabilities of both individual and collective team members.

\section{Result and Discussion}\label{sec2}

\begin{figure*}[t]
	\centering
	\includegraphics[width=1.0\linewidth]{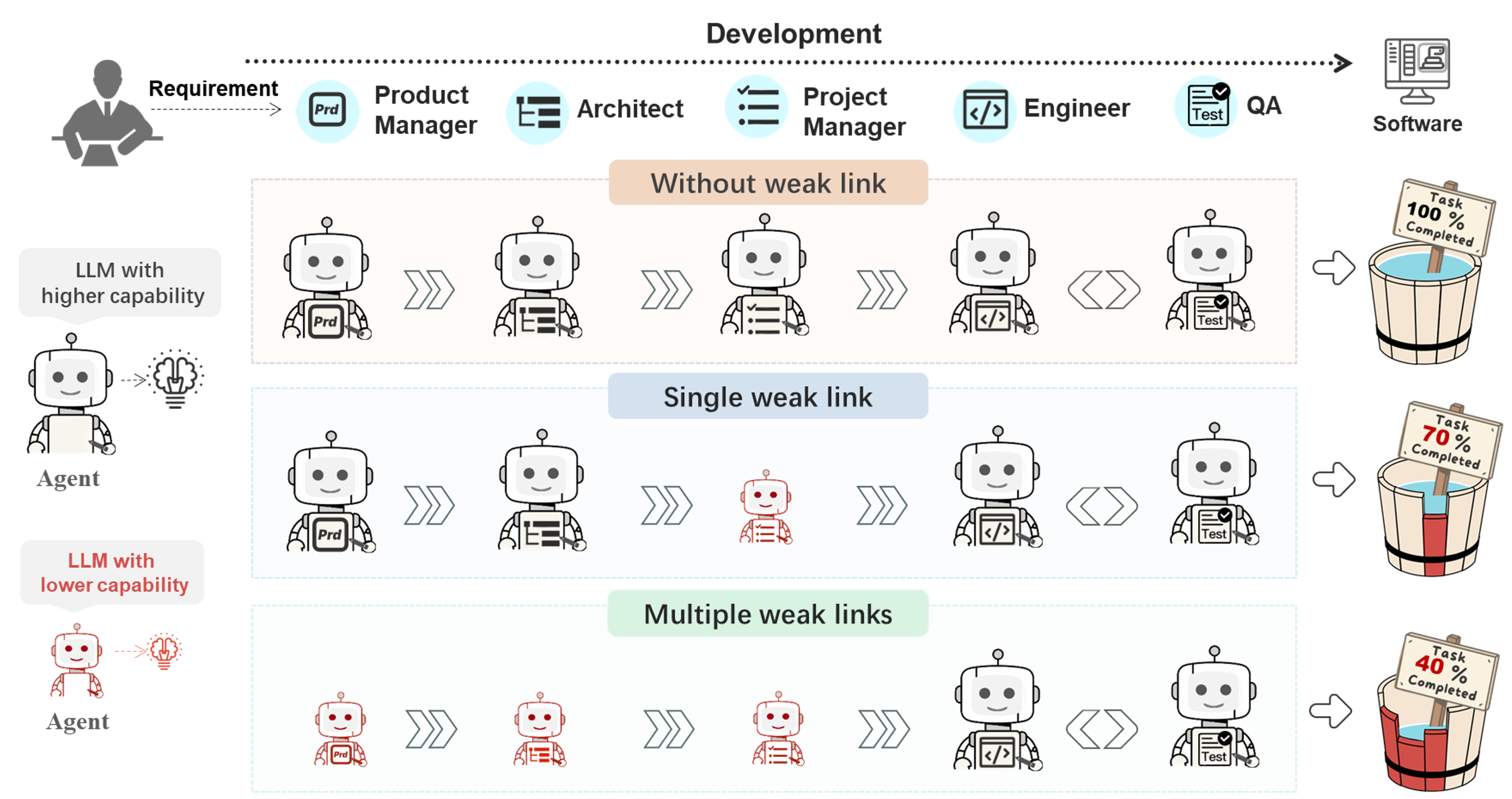}
	\caption{Schematic diagram of virtual team and the study design. There are three types of team configures are employed. (i) Homogeneous team. All team members are driven by the same LLM. (ii) Single weak link. One team member is replaced using an LLM with lower capability. (iii) Multiple weak links. Multiple team members are substituted for those driven by LLMs with lower capabilities.}
	\label{fig:1}
\end{figure*}

In this paper, we construct a virtual software development team consisting of five members: product manager (PRM), architect (ARC), project manager (PJM), engineer (ENG), and quality assurance engineer (QA), who are responsible for software requirement analysis, system architecture design, development scheduling, code writing, and testing/debugging, respectively.
The details of the virtual team configuration are described in Materials and Methods.
Software development is a classic example of multi-process problem in engineering management,
involving not only coding but also requirements analysis, development documentation, architectural design, testing, and  collaboration \citep{scholtes2016aristotle,betti2025dynamics}.
The configuration of the virtual team is detailed under Materials and Methods.

As demonstrated in Figure~\ref{fig:1}, the virtual development team has been configured with a specific architecture.
We focus on the analysis of team performance under three distinct team configures to quantifying the impact of weak links on overall performance:
\begin{enumerate}
	\item Homogeneous Team: To investigate the collaborative performance affected by the commom capability of LLM, all team members are driven by the identical LLM;
	\item Heterogeneous Team of Single Weak Link: To examine the impact of a single weak link on team performance, only one team member is driven by an LLM with lower capabilities;
    \item Heterogeneous Team of Multiple Weak Links: To quantify the cumulative effect of multiple weak links on team performance, each team member driven by an LLM with lower capabilities are sequentially introduced.
\end{enumerate}

\subsection{Homogeneous team}
\begin{figure*}[htb]
	\centering
	\includegraphics[width=1.0\linewidth]{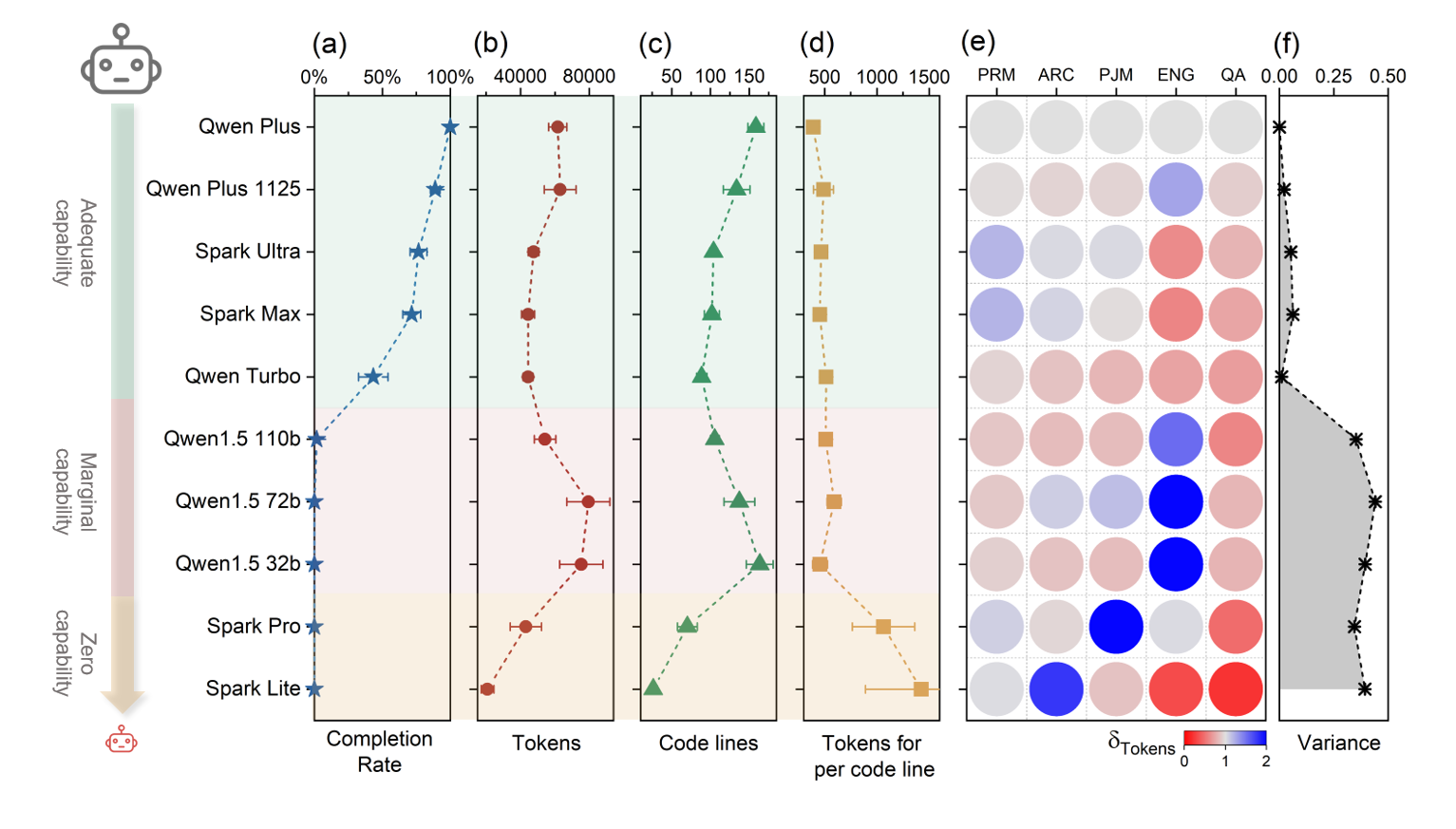}
	\caption{The team performances of the homogeneous team for different capabilities. (a) The completion rate. (b) The total token consumptions of the teams. (c) The final number
		of code lines. (d) The tokens for per code line. Error bars in (a)-(d) correspond to the standard error across ten experiments. According to these metrics, the teams driven by different LLMs are divided into three states: adequate capability, marginal capability and zero capability. (e) The normalized token consumption for each team member and (f) denotes the variances.}
	\label{fig:2}
\end{figure*}

In the homogeneous case, we configure team members using the same LLM setup and explore the boundaries of team capabilities in completing specific tasks by employing different LLMs.
As presented in Figure 2(a), we apply ten distinct LLMs to power the agents.
These include Qwen Plus, Qwen Plus 1125, Qwen Turbo, Qwen1.5 110b, Qwen1.5 72b, Qwen1.5 32b, Spark Ultra, Spark Max, Spark Pro, and Spark Lite from the LLM families of Qwen (Alibaba Cloud, API Version\citep{Qwen}) and iFlytek Spark (iFlytek Cloud, API Version \citep{iFlytek}), respectively.
To mitigate the impact of randomness in single simulations, each group experiment is carried out ten times.
To evaluate completion status, we define six function points to calculate the project completion rate as detailed under Materials and Methods.
As shown in Figure~\ref{fig:2}(a), the ranking of team capabilities is based on the completion rates of task, with teams achieving a completion rate of 0\% being ranked according to the number of tokens consumed.
As LLM communication and development are text-based, the workload of LLM-driven agents can be measured by the number of tokens utilized.
The total token consumptions of teams with different capabilities are illustrated in Figure~\ref{fig:2}(b).
In the context of software development projects, the number of lines of code (LOC) is a critical metric for evaluating team output, as presented in Figure~\ref{fig:2}(c).
Consequently, the tokens for each code line are defined as a measure of team efficiency.
A lower value indicates higher development efficiency, as demonstrated in Figure~\ref{fig:2}(d).

Based on the teams' performances, three distinct states of team capability in completing projects are identified: \textit{adequate capability}, \textit{marginal capability}, and \textit{zero capability}.
When overall capability is high, the team is referred to as in a \textit{adequate capability} state, which signifies the ability to complete several function points, albeit with variable completion rates.
As team capability is sufficiently strengthened, the project completion rate attains 100\%.
However, as capability diminishes, project completion rates decline.
This is due to the limited capacity of the team, which results in lower workloads and code volumes.
As a consequence, the completion of all function points is prevented.
As demonstrated in Figure~\ref{fig:2}(d), within this regime, the collective efficiency of the team remains consistently high.
Despite the increase in the development of code lines and workload, the efficiency of the team has been shown to improve with enhanced capability.

With further reductions in capability, the team has become entirely incapable of completing tasks, achieving a task completion rate of approximately zero.
However, two states emerge in this scenario: marginal capability, and zero capability.
Notably, in the \textit{marginal capability} state, which occurs near the completion threshold, an intriguing phenomenon emerges.
The team generates significant amounts of codes and tokens, but fails to complete tasks, resulting in a substantial waste of effort.
Within this regime, while team efficiency remains high, it falls into the pseudo-efficiency trap.
In this context, the observed behavior of the team is reminiscent of the mythological figure of Sisyphus, endlessly endeavoring to achieve a goal that remains unattainable due to an absence of the necessary capabilities, and repeatedly producing work deemed to be ineffective.
We term this as "\textit{Sisyphus Predicament}".

As capability continues to decline, the gap between capability and goals widens, eventually reaching the \textit{zero capability} state.
At this point, the team abandons project development, achieving only the basic framework while leaving core content unfinished.
Consequently, both code lines and workloads are minimal, however, team efficiency becomes notably low.
This demonstrates that when the capability is significantly mismatched with the goal, not only are tasks uncompleted, but efficiency also becomes exceedingly low.

The results emphasise the significance of effective team management, particularly in circumstances where teams encounter a Sisyphus Predicament.
The existence of pseudo-efficiency trap during project execution indicates that short-term performance metrics alone may not reflect true team capability, with project failures often surfacing only at the end.
This suggests that team leader should conduct regular assessments of team performance, especially when workloads are high,
and make timely adjustments to team composition and resource allocation.
This is to be done to avoid the unproductive cycle of a Sisyphus Predicament and resource wastage.

In order to investigate the underlying mechanism for different states of team performance, we carried out a systematic analysis of the workload distribution for each team member.
The distribution of work among the members of the Qwen Plus configuration team, which has a 100\% completion rate, is considered to be a reasonable allocation of accomplishing the task.
Utilising this established benchmark, we then normalize the token consumptions $\delta_{tokens}$ of members across teams with different capabilities.
As shown in Figure~\ref{fig:2}(e), in the marginal and zero capability regimes, where the completion rate is 0,
the normalized workload of team members exhibit significant imbalance.
Some members showing notably higher completion levels relative to the reference value, while others are notably lower.
It is evident that the workload of some members is significantly higher than the benchmark, while that of others is considerably lower.
This imbalance reveals the presence of an unreasonable distribution of workload within the team.
In contrast, in the adequate capability regime, the workload of team members shows small fluctuations, remaining close to the value of benchmark.
In order to quantify the observed discrepancy in distribution, the variance of the normalized workloads for the five team members is calculated, as illustrated in Figure~\ref{fig:2}(f).
The result further validated our observation that there is marked variability in the workload distribution among members in the zero percentage completion rate regime.
This finding suggests that unreasonable task distribution may be a critical factor contributing to project failure, necessitating close monitoring during project execution.

\subsection{Single weak link}

The heterogeneity of team composition significantly impacts team performance, particularly when team members possess varying levels of capability \citep{chiang2024chatbot}.
A critical question to be addressed is how much impact a weak link within a team can have on the entire group.
We select Qwen Plus and Qwen1.5 32b as representative models for strong and weak capability, respectively.
By sequentially replacing each member in the homogeneous group with a weak-capability agent, we systematically quantify the contribution of each member to the team performance.

Team performance is evaluated based on three key metrics:
1. \textit{Efficiency}. Defined as the amount of code lines produced per thousand tokens consumed, reflecting output per unit of labor.
2. \textit{Productivity Ratio}. Measured as the ratio of the project completion rate to the cost, where the cost is calculated based on the price per token for each LLM.
Detailed pricing is available in Supplementary material, Table~S3.
This represents the unit cost of the investment in completing effective work on the project.
3. \textit{Completion Rate}. A critical indicator of project quality.
These metrics evaluate team performance from three perspectives, including efficiency, cost, and quality, respectively.
A homogeneous group in which all members are driven by Qwen Plus serves as the benchmark.
In this framework, a higher score across all three metrics indicates superior team performance.

\begin{figure*}[htb]
	\centering
	\includegraphics[width=0.96\linewidth]{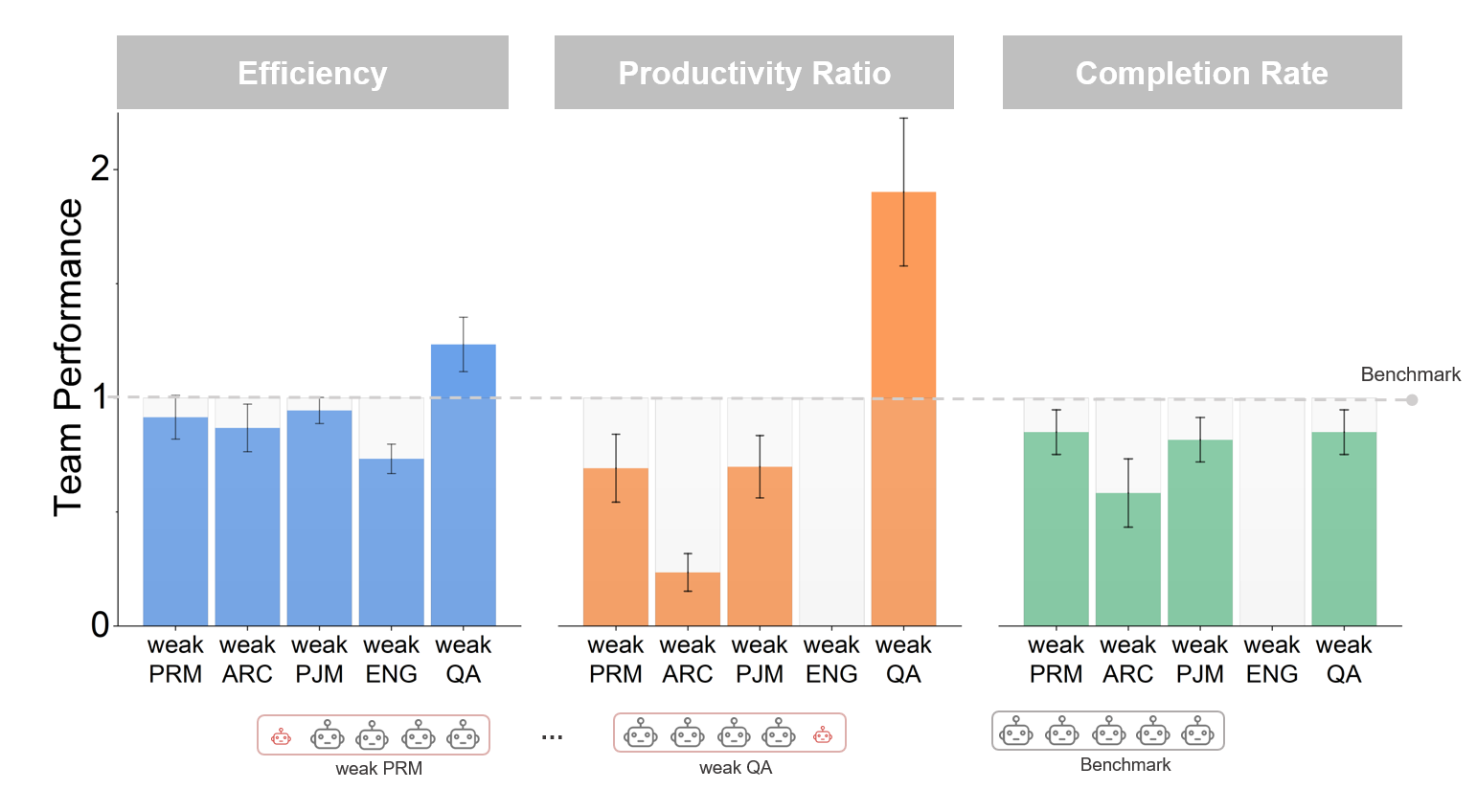}
	\caption{The short-board effect for each team member analyzed from three perspectives: efficiency, productivity ratio, and completion rate. The dashed line shows the benchmark which is the team performance of the homogeneous team. Error bars correspond to the standard error.}
	\label{fig:3}
\end{figure*}

As shown in Figure~\ref{fig:3}, upon the integration of the weak links formed by the five members, a benchmark comparison of team performance is conducted across three critical metrics.
The analysis revealed significant variations in how different members affect team performance.
Specifically, a weak ARC results in a minor reduction in team efficiency.
Yet an unreasonable architecture can lead to a substantial increase in overall costs and a significant impact on project completion rates.
In such a development project, the ENG is core member, the other members are non-core members.
The results reveal that a weak ENG not only directly causes project failure, but also plunges the team into the Sisyphus Predicament, characterised by pseudo-high efficiency and significant ineffective efforts.
In contrast, the weakening of the non-core members such as PRM and PJM, leads to varying degrees of decline in team performance.
Interestingly, although a weak QA reduces completion rate, it paradoxically improves efficiency and reduces costs, demonstrating an inverse short-board effect.
This is primarily because the weak QA reduces interactions with ENG, even though it fails to accurately identify project defects, it significantly reduces the debugging workloads of the ENG.

The results clearly show that, although weak links in different members may affect team performance in different ways, weakening any member has a negative impact on the project quantity.
Specifically, the core member plays a decisive role in project outcomes.
If this member is weak, it will trap the team in a Sisyphus Predicament and deal a devastating blow to project progress.
Yet, this issue cannot be timely detected through efficiency assessments alone during execution.
Although functional non-core members do not directly cause project failure, they still significantly affect quantity, efficiency, and cost.
This underscores the necessity for manager to conduct comprehensive assessment and optimise team configuration across all dimensions.

\subsection{Multiple weak links}
We quantitatively calculated the impact on team performance caused by each member forming a weak link across three dimensions: efficiency, cost, and quality.
This prompts an intriguing question: when a team contains multiple weak links, does the commonly held belief that a team is only as strong as its weakest link become invalid?
Consequently, we methodically introduce multiple weak links into the software development team, with the objective of evaluating the performance of the team.
The team driven by Qwen Plus model is still utilised as the benchmark,
with agents driven by the lower-capacity Qwen1.5-32b model being introduced incrementally starting from the first member.

\begin{figure*}[htbp]
	\centering
	\includegraphics[width=0.9\linewidth]{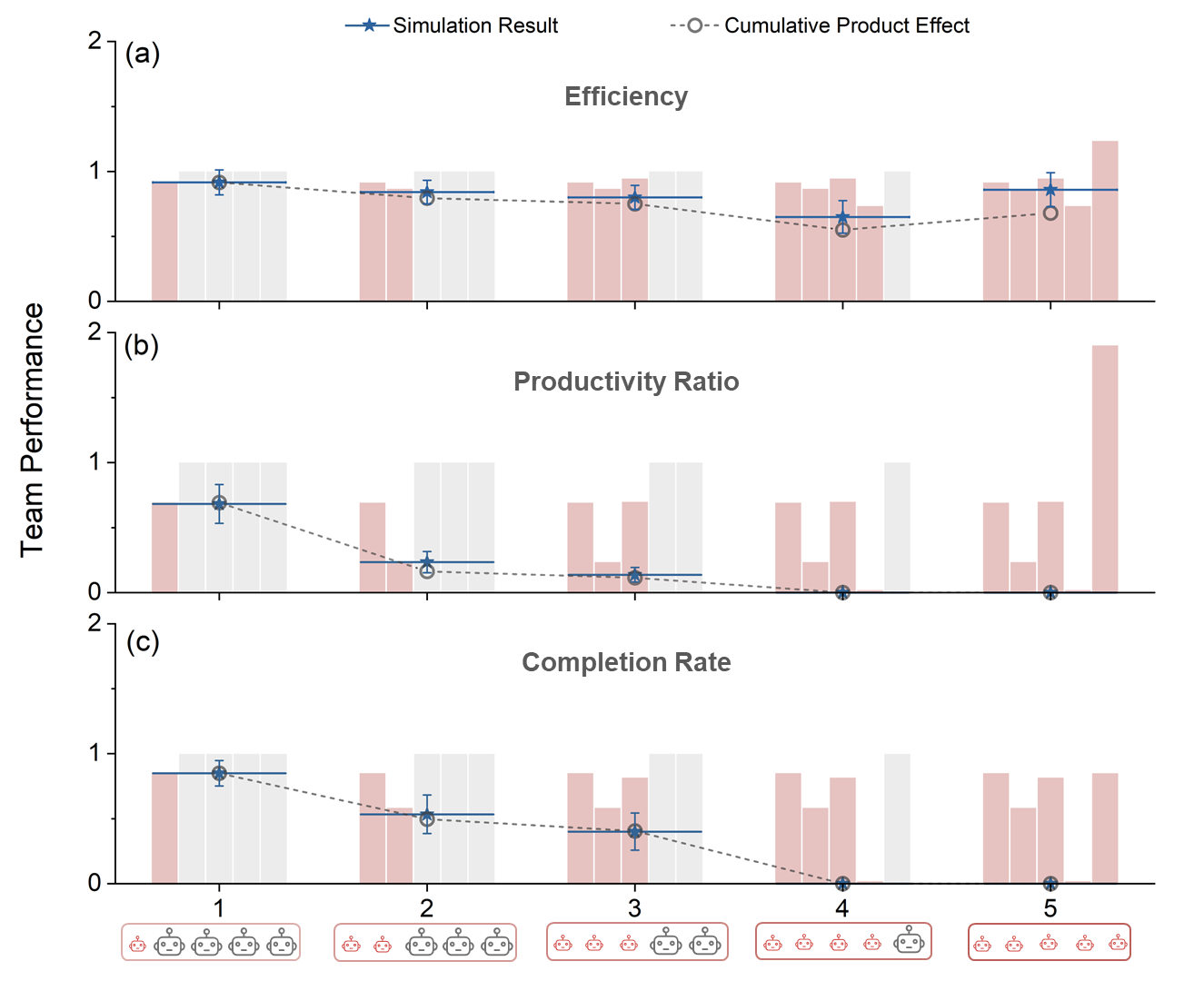}
	\caption{The simulated team performance and cumulative product effect for the team with multiple weak links. The background bars represent the team configurations. The red bars and grey bars denote the members with and without weak links, respectively. The stars with lines are the simulation results, and the error bars correspond to the standard error. The hollow circles show the cumulative product effect.}
	\label{fig:4}
\end{figure*}

As illustrated in Figure~\ref{fig:4}, the background bars represent the impact of individually introducing this member as weak link on team performance,
as obtained in the above section, reflecting the configuration of the team.
The performance of the team is evaluated using three metrics.
The blue stars indicate the simulated results of team performance under each configuration.
One can be observed that, with the exception of the final data point in Figure~\ref{fig:4}(a),
the performance across all three aspects for the remaining team configurations is lower than the impact caused by any single weak link in the team.
This result suggests that the collective performance of a team is not dominated by its weakest link,
but rather, when multiple weak links are present, the overall performance is constrained to being less than that of the weakest link.

Upon analysis, we find that the team performance is determined by the cumulative product effect of multiple weak links.
Assuming the impact in team performance caused by a single member $m_i$ is $P(m_i)$, the overall team performance can be expressed as $P(team)$ = $\prod_{m_i} P(m_i)$.
The results of the cumulative product effect, represented by the black hollow circles in Figure~\ref{fig:4},
agree well with the simulated results of each configuration, confirming the validity of the cumulative product effect.
For example, in terms of important completion rate, even if the core rigid demand member ENG is highly capable,
when the functional non-core members (PRM, ARC, PJM) are weak, the team overall completion rate declines significantly.

As demonstrated in the preceding section, a weak QA has been shown to have a positive impact on team efficiency.
An interesting phenomenon can be observed in the result is that this enhancing effect remains valid in the cumulative product effect.
The decline in team performance caused by multiple weak links is partially compensated by the final QA, leading to an improvement in efficiency as demonstrated in the final data point of Figure~\ref{fig:4}(a).
Such phenomenon is also predicted by the cumulative product effect.
However, for productivity ratio and completion rate, since the team becomes unable to complete tasks (i.e., P(ENG) = 0), the QA effect is invalidated in the cumulative product effect.
We also conducted studies using other LLMs as benchmarks and weak links, and the results are consistent with the above findings.
The related results are shown in Supplementary material, Figure~S1-S4.

The cumulative product effect is an significant finding that not only illustrates the quantitative relationship between a collective team performance and the capabilities of the individual members,
but also serves as a stark caution against overlooking multiple weak links within a team.
Although the impact of a single weakness may be negligible, the accumulation of multiple weak links can ultimately result in catastrophic failure for the entire team.
Furthermore, this also provides a distinct insightful guidance for team management.
In traditional project management, it is commonly believed that improving team performance requires strengthening the weak links.
However, our results indicate that when weak links cannot be easily improved, enhancing the capabilities of other team members may serve to mitigate the impact of these weak links.
This assertion is contrary to the hypothesis that a team is as strong as its weakest link.
Such insight is not only relevant to management but may also be applicable to other fields such as education, personal development, scientific discovery, and even broader realms.

\section{Conclusion}
In this paper, we employ LLM-driven agents to study the impact of individual capability on team performance in collaborative environments.
Three distinct states of relationship between team performance and capability are identified in homogeneous team configurations.
Notably, when capabilities hover at critical thresholds, the team falls into a "Sisyphus Predicament" state, characterized by extensive ineffective efforts and pseudo-high efficiency.
An analysis of team members reveals that the failure of the team arises from mismatched capability and objective, leading to unreasonable workload distribution.
Furthermore, we develop a framework to quantify the short-board effect through introducing a single weak link into a high-capability team.
The different impacts of weak links across members on overall team performance are assessed.
In addition to confirming the pivotal function of core member, the results accentuate the crucial contribution of non-core members,
where such inadequacies can exert a detrimental effect on performance in terms of efficiency, cost, and quality.
Interestingly, there is an anti-short-board effect, in which the weak link paradoxically enhanced team efficiency.
More importantly, a cumulative product effect emerges from the introducing of of multiple weak links. 
The performance of a team is not solely determined by the weakest component, but rather by the cumulative interactions among multiple weak links.

The anti-short-board effect likewise accumulates, suggesting that it is feasible to enhance team performance through strengthening weak links.
Thus, targeted improvements in non-weakest members can partially compensate for the adverse effects.
These findings enrich theoretical frameworks and yield substantial implications for optimizing team management, organizational operations, supply chain resilience. 

\section{Methods}\label{sec11}

\subsection{Virtual team configuration}
The virtual development team is constructed based on the MetaGPT framework \citep{hong2024metagpt} and comprises five agents: product manager, architect, project manager, code engineer, and quality assurance engineer.
Descriptions of team members and the prompt of the develop requirement are presented in the Supplementary material, Table~S1.
Team members adhere to standardized operating procedures, proceeding sequentially according to their roles in the development process.
The code engineer and quality assurance engineer engage in iterative communication to facilitate coding and debugging.
To prevent potential infinite loops, the testing process is capped at a maximum of five rounds.
In this study, the development of a classic Snake game is selected as the demand scenario.
This is a scenario which has a clear objective and is conducive to quantitative evaluation of the development outcome.
Although this task is relatively simple, the best-performing individual LLM fails to achieve a 100\% completion rate without adopting a team-based approach, reaching only 95\%. 
Moreover, the performance gap between individual LLMs and team configurations persists across other model variants, suggesting that collaborative dynamics play a non-negligible role in a standard operating procedure with a sequential team. 
The simplicity of the task further enables us to focus on team-level behaviors, minimizing confounding effects from task complexity.

\subsection{Metrics of team performance}
To assess team performance across multiple dimensions, we quantified various outputs.
Given that LLMs operate through text-based communications, workload is measured by the number of tokens consumed, denoted as $T$.
The number of lines of code in the final developed software serves as a key metric of team output, denoted as $N_{loc}$.
For software quality, We established six function points as evaluation criteria for software completion, with detailed requirement descriptions and function point definitions provided in Supplementary material, Table~S2,
with each completed point contributing $1/6$ to the project completion rate $R$.
The cost of the project is calculated based on the per-million-token pricing of LLMs.
The pricing details for the models employed are provided in Supplementary material, Table~S3, primarily referred to the model capabilities and official API.
Total project cost is given by $C = \Sigma T_{m_i}  C_{m_i}$, where $T_{m_i}$ and $C_{m_i}$ represent the tokens consumed by the member ${m_i}$ and the corresponding pricing, respectively.
Building on this, we define other two additional performance metrics: efficiency, which is the workload per line of code developed, $E = \left\langle T / N_{loc} \right\rangle $, and productivity ratio $Pr = \left\langle R / C \right\rangle $,
where $ \left\langle \dots \right\rangle $ denotes the sample average.
For each team configuration, we perform around ten independent simulations.
These metrics evaluate team efficiency and effective cost, respectively.
In team configurations involving weak links, following the selection of a benchmark,
the normalized team performances are expressed as $\bar{E} = E / E_{benchmark}$, $\bar{Pr} = Pr / Pr_{benchmark}$, and $\bar{R} = R / R_{benchmark}$.
In analyses of the cumulative product effect, all team performances $P_{m_i}$ are employed represent normalized team performance.

\section*{CRediT authorship contribution statement}
\textbf{Xin Xu:} Conceptualization, Formal analysis, Methodology, Visualization, Writing – original draft, Writing – review \& editing. \textbf{Jiu Zhang:} Methodology, Investigation, Formal analysis, Visualization, Writing – review \& editing. \textbf{Xiao-Ling Lei:} Supervision, Conceptualization. \textbf{Xiong-Fei Jiang:} Conceptualization, Methodology, Formal analysis, Investigation, Writing – review \& editing. \textbf{Long Xiong:} Conceptualization, Formal analysis, Methodology, Software, Visualization, Project administration, Investigation, Writing – original draft, Writing – review \& editing.

\section*{Declaration of Competing Interest}
The authors declare that they have no known competing financial interests or personal relationships that could have appeared to influence the work reported in this paper.

\section*{Acknowledgements}
This work was supported in part by National Natural Science Foundation of China under Grant No. 12305053, Yunnan Fundamental Research Project under Grant No. 202401CF070167, High-level Scientific Research Incubation Project of Ningbo University of Finance \& Economics under Grant No. 1320263705, Natural Science Foundation of Fujian NO.2026J008332.


\bibliography{sample}

\end{document}